# The Cosmic Microwave Background Radiation Power Spectrum as a Random Bit Generator for Symmetric- and Asymmetric-Key Cryptography

Jeffrey S. Lee[1]
Gerald B. Cleaver[1,2]

[1]Early Universe Cosmology and Strings Group
Center for Astrophysics, Space Physics, and Engineering Research
[2]Department of Physics
Baylor University
One Bear Place
Waco, TX 76706

Jeff_Lee@Baylor.edu
Gerald_Cleaver@Baylor.edu

*In this note, the Cosmic Microwave Background (CMB) Radiation is shown to be capable of functioning as a Random Bit Generator, and constitutes an effectively infinite supply of truly random one-time pad values of arbitrary length. It is further argued that the CMB power spectrum potentially conforms to the FIPS 140-2 standard. Additionally, its applicability to the generation of a (n×n) random key matrix for a Vernam cipher is established.*

______________________________________________________________________

## 1. Overview of the CMB

A remnant of the Epoch of Recombination in Big Bang cosmology, the CMB is a pervasive, highly spatially isotropic radiation field permeating all of the visible universe. With a thermal blackbody temperature of 2.72548 ± 0.00057 K [1], the spectral radiance peak occurs at 1063.3 $\mu$m (282.1 GHz).

The power spectrum of the CMB has been well-mapped by the *Planck* mission[1], and the likelihood has been determined by the Planck data and described in detail by Ade et al. [2]. The CMB power spectrum is a unique spectral fingerprint of the core cosmological model. *Planck*'s map, made over a 4π sr solid angle and to a precision of several arcminutes, results in the power spectrum being attainable over an unprecedented range. The estimation of cosmological parameters from the power spectrum necessitates an uncertainty-propagating likelihood function, a property which can also be particularly useful for cryptographic purposes. At low multipoles, an approach based on Gibbs sampling is applicable [3], [4] because the power spectrum is non-Gaussian at large scales. For high multipoles, a pseudo-$C_\ell$ technique is useful [5].

In the case of high multipoles, the *Planck* maps are comprised of $5 \times 10^7$ pixels per detector, and the determination of a per pixel likelihood would be computationally impractical. However, implementation of a data compression algorithm in the form of pseudo-$C_\ell$ power spectra would result in a significant reduction in processing times and negligible information loss.

---

[1] *Planck* (http://www.esa.int/Planck) is a European Space Agency (ESA) project carrying instruments supplied by, in particular, France and Italy. Contributions were made by NASA, and telescope reflectors were provided by an ESA-Denmark-led scientific consortium.

In contrast to the CamSpec likelihood, in which the deconvolved spectra is formed without any prior smoothing of the pseudo-spectra $\tilde{C}_l^{ij}$ (given by eqn. (2)), the $P_{\text{lik}}$ likelihood, which is often used for robustness and cross-checks tests [6], is chosen.

$$\tilde{C}_l^{ij} = \frac{1}{2l+1} \sum_m \tilde{T}_{lm}^i \tilde{T}_{lm}^{j\dagger} \qquad (1)$$

where $\tilde{T}_{lm}^i$ are the spherical harmonic coefficients of the weighted temperature map as produced by detector $i$; $l$ is the pseudo-spectrum multipole; $(i, j)$ is the detector pair, and † indicates the Hermitian transpose.

The exact likelihood for a full-sky Gaussian signal is given by:

$$p(\text{maps} \mid \theta) \prec \exp\left[-\sum_l (2l+1)\kappa\left(\hat{C}_l, C_l(\theta)\right)\right] \qquad (2)$$

where $\theta$ is the signal model parameters vector; $\hat{C}_l$ are the empirical angular spectra, and $\kappa\left(\hat{C}_l, C_l(\theta)\right)$ is the Kullback divergence between $n$-variant Gaussian distributions with zero-means and covariant matrices $\hat{C}_l$ and $C_l(\theta)$. $\kappa\left(\hat{C}_l, C_l(\theta)\right)$ is given by [2]:

$$\kappa\left(\hat{C}_l, C_l(\theta)\right) = \frac{1}{2}\left[\text{tr}\left(\hat{C}_l, C_l(\theta)^{-1}\right) - \log \det\left(\hat{C}_l, C_l(\theta)^{-1}\right) - n\right] \qquad (3)$$

Performing cutoffs of the observable sky introduces off-diagonal terms of the covariance between various monopoles. However, with clever binning, the off-diagonal terms are insignificant, and slowly varying spectra sources, such as the CMB anisotropies, can be modeled. In so doing, eqn. (2) becomes

$$p(\text{maps} \mid \theta) \prec \exp[-L(\theta)] \qquad (4)$$

where $L(\theta) = \sum_{r=1}^R n_r \kappa\left(\hat{C}_r, C_r\right)$.

In this case, the angular spectra $C_l$ for each of the cross-frequency spectra have been averaged and placed in $R$ spectral bins. The associated spectral windows $w_r(l)$ with $r = 1...R$ result in:

$$\hat{C}_r = \sum_l w_r(l)_l \hat{C}_l, \quad C_r = \sum_l w_r(l) C_l \qquad (5)$$

where $w_r(l)$ is the window function for the $r^{th}$ bin. $C$ represents both binned and unbinned spectra.

If the criteria for spectral binning are

$$w_r(l) = \begin{cases} \dfrac{l(l+1)(2l+1)}{\sum_{l_{min}^r}^{l_{max}^r} l(l+1)(2l+1)} & l_{min}^r \leq l \leq l_{max}^r \\ 0 & \text{otherwise} \end{cases}, \quad (6)$$

then the effective number of modes in the $r^{th}$ bin is

$$n_r = f \frac{\left(\sum_l w_r(l)^2\right)^2 (2l+1)}{\sum_l w_r(l)^4} \quad (7)$$

where $f$ is the observed sky frequency.

This binned likelihood is in particularly good agreement with the primary likelihood, even though all couplings between the various multipoles are not captured; the effective computational speed is greatly increased. This rise in computational speed makes performing a vast set of robustness tests possible because the number of tested parameters per unit time is significantly boosted. Additionally, the investigation of instrumental effects, such as spurious signals in the electronics of the detectors, can be expedited swiftly, and the agreement between detector pairs within a frequency channel can be promptly ascertained [2].

Since the $P_{lik}$ likelihood includes the empirical auto-spectra, it allows for a direct estimate of the detector noise power spectra. Additionally, the noise estimates agree well with the noise spectra employed in the formation of the CamSpec likelihood covariance matrix [2]. To produce the CMB power spectrum, a two-step process is employed. First, all parameters, including the noise, are estimated using both the auto- and cross-spectra. Second, the noise estimates are fixed, and a fiducial Gaussian approximation allows for the exploration (and ultimately the extraction) of the residual free parameters, including the CMB power spectrum and excluding the auto-spectra [2]. The extracted CMB power spectrum can then be used as a private encryption key or to generate an asymmetric key pair.

## 2. The Cryptographic Application

The effect of key length on cryptographic security has been well established [7], [8], [9], [10], [11], [12]. The CMB power spectrum, the binning and extraction of which is described in the previous section, is an unpredictable, arbitrarily large and totally random number[2] that can be used by Alice (a sender) to generate an acceptable public- and private-key pair appropriate for use by an asymmetric key algorithm (see Figure 1).

---

[2] Or a matrix, depending on whether detector pairs are considered.

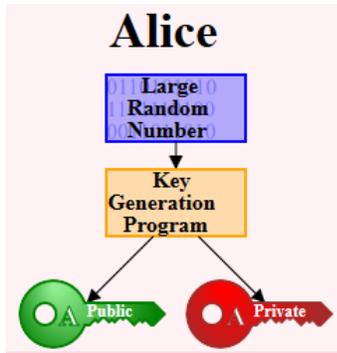

Figure 1: Diagrammatic representation of the use of a large random number to implement a key generation program[3].

Alice's acquisition of the CMB power spectrum at any chosen sky frequency can be done with arbitrary precision (to the limit of the sensitivity and subject to the signal to noise ratio of the detectors) and for an arbitrary period of time *appears to be* a strategy that could also be employed by Eve (an adversary). Of course, this presupposes that Eve would have gained prior knowledge of the stellar coordinates from which CMB power spectrum was observed, the frequency of observation, the time and duration of the observation(s), the binning parameters, etc.

However, even allowing for the possibility that Eve was privy to such specific information, the exact duplication of the Alice's random number generated from the CMB power spectrum is not possible due to random variations in photon energy at any sky frequency, spurious signals within the detectors, interference from other sources of stellar radio noise, etc. Therefore, the digitized CMB power spectrum obtained by Alice is unique and cannot be acquired through "identical" power spectrum observations of the CMB by Eve.

It is noteworthy that although a similar encryption strategy could be (and has been) implemented with irrational numbers, the digits of which would provide true randomness and an arbitrarily large key space, the number, once known, is rendered a one-time pad. Repeated CMB power spectrum measurements would allow the same key space source to be used indefinitely, and knowledge of the origin of the key space source would not nullify the CMB power spectrum as a key space source (i.e., if a key space source irrational number is discovered by Eve, then the pattern of digits is known by Eve; this would not be true of the CMB power spectrum). Of course, there is no shortage of available irrational numbers; however, even Alice does not know the digits of the CMB power spectrum key space prior to measurement. Once the CMB power spectrum has been resolved, this information would need to be transmitted to Bob (a receiver) by one of the conventional methods, making it neither more nor less secure than any other key transmitted the same way. Such an approach would be arguably effective for symmetric-key algorithms. In the case of asymmetric-key algorithms, the private key could be generated by means of the CMB power spectrum, and the private key would obviously need be created with parameters by which its calculation, by Eve, would be computationally infeasible.

While all-encompassing of the visible universe and well-studied, it's important to point out that the CMB is by no means the only truly random astrophysical Random Bit Generator (RBG). The 21 cm hydrogen line, planetary (non-terrestrial) radio noise, supernova remnants, radio galaxies, etc. are all viable candidate RBGs. None of the above-mentioned potential RBGs require the deployment of interplanetary spacecraft or even satellites. A terrestrial radio telescope would suffice, thus making astrophysical entropy sources accessible on comparatively modest budgets.

---

[3] Image from: https://upload.wikimedia.org/wikipedia/commons/3/32/Public-key-crypto-1.svg

## 3. The CMB Power Spectrum RBG's Potential for Conformity to FIPS 140-2

Argued here is that the CMB power spectrum method of random number generation to produce both private- and public-keys meets the requirements for conformity to the United States government computer security standard FIPS 140-2 (Federal Information Processing Standard) used for the accreditation of cryptographic modules as laid out in the United States Department of Commerce (DOC) Recommendations [13]. The DOC Recommendations advise that all encryption keys be based directly or indirectly on the output of an approved Random Bit Generator.

Keys indirectly generated from a RBG include those derived during an agreement transaction [14], derived from another key using a key derivation function [15], or derived from a password for storage applications [16] because an ancestor key[4] or random value[5] was obtained directly from the output of an approved RBG. Although deconvolution of a CMB power spectrum is not currently specified as an approved RBG, its potential candidacy for approval is strong; the same argument can be made for other astrophysical sources of entropy.

The two sets of cryptographic algorithms necessitating cryptographic keys, which have been approved for use by the United States government, are asymmetric-key algorithms and symmetric-key algorithms, for which random number generation from a CMB power spectrum is viable.

Additionally, the RGB or portion of the RGB cryptographic module [17] that generates the key must "reside" within the FIPS 140 key-generating module[6]. In the case of the CMB power spectrum, the entropy source clearly does not co-reside with the random-output-generating algorithm. The transporting of generated keys using secure channels and their use by the associated cryptographic algorithm within FIPS 140-compliant cryptographic modules are implied, irrespective of the random-number-generating entropy source.

The security strength of the RGB must be adequate for the protection of the target data. Details on the security strength for approved RGBs can be found in [17], [18], [19]. A key's security strength is dependent on the key size [20], the key-generating process, the security strength of the method of transporting the key, and of course, the algorithm by which it was generated.

It is notable that the supported security strength by a key is not dependent exclusively on the length of the key. For instance, if a $n$-bit key is used with an AES-$n$ (e.g., AES-128) for the encryption of plaintext, then an AES operation employing that key can provide only $m$ bits of security strength if $m < n$. While the key strength is often less important to the security of a cryptographic cipher than the strength of the algorithm(s), a strong key is never disadvantageous.

### 3.1 RBG Output

The DOC Recommendations require that approved output of a RBG boast adequate security strength to protect the plaintext by means of a symmetric key or the needed random number for the generation of an asymmetric key pair (for which an approved generating algorithm is required).

---

[4] A key used to generate another key.
[5] For example, a random value that was used to create a key-agreement key pair.
[6] The cryptographic module in which the key is generated.

For key/random number generation, the following criterion is imposed:

$$K = V \oplus W \qquad (8)$$

where $K$ is the key generated by an approved RBG (symmetric key) or a random number generated by an approved RBG (asymmetric key pair); $W$ is the required length bit string with the capability of protecting the plaintext, and $V$ is bit string equal in length to $W$ and determined in a fashion that is independent of $W$.

The lack of restrictions placed on the choice of $V$ requires that the means by which $W$ is selected bestow the crucial entropy. The selection of $W$ and/or $V$ from a CMB power spectrum will allow for any required bit string length to protect plaintext of any size and importance. The computational and informational independence of $V$ and $W$ assures that neither bit string can be determined from the other. Clearly, the implementation of a CMB power spectrum as a RBG will not violate this requisite.

## 4. The Application of the CMB Power Spectrum to a Vernam Cipher

The CMB power spectrum has a direct application to symmetric key cryptography and the creation of a Vernam cipher[7], in that it can serve as a key for the generation of a random key matrix. In their paper, Nath et al. consider a reasonably robust (16×16) random key matrix, and make use of a 16-character key, which could be comprised of any of the 0-255 ASCII characters [21]. Although significantly more computationally intensive, the CMB power spectrum allows for a ($n \times n$) random key matrix, and each "character" would be represented by a string of $m < n$ digits. From the data in Table 1, Nath et al. generate their key with eqn. (9) [21]:

| Length of Key ($n$) | 1 | 2 | 3 | 4 | 5 | 6 | 7 | 8 | 9 | 10 | 11 | 12 | 13 | 14 | 15 | 16 |
|---|---|---|---|---|---|---|---|---|---|---|---|---|---|---|---|---|
| Base ($b$) | 17 | 16 | 15 | 14 | 13 | 12 | 11 | 10 | 9 | 8 | 7 | 6 | 5 | 4 | 3 | 2 |

Table 1: Key length and base for a 16×16 random key matrix **[21]**.

$$S = \sum_{i=1}^{n} C b^i \qquad (9)$$

where $S$ is the key sum; $n$ is the length of the key; $C$ is the ASCII code, and $b$ is the base.

Implementation of a CMB power spectrum using this method would require that $C$ in eqn. (9) be replaced by a character code comprised of either:

1. a series of random digits in the intensity measure of the resolved CMB power spectrum with known measurement parameters arriving at a single detector, as discussed in Section 1 (i.e., sky frequency, binning parameters, spatial coordinates, time of measurement, duration of measurement, etc.), or
2. a series of random digits from multiple intensity measurements of the resolved CMB power spectrum arriving at multiple detectors and with known measurement parameters (as in case #1).

---

[7] Also called a "one-time pad".

This modification to the method of Nath et al. allows for a much larger key space, the encryption and decryption of which can be computationally feasible, but for which brute force attacks upon the cipher become exponentially infeasible. Furthermore, the CMB power spectrum generated key space can be constructed to be significantly in excess of the computing capabilities of current hardware architecture, thus imposing upper limits on the practical key space size.

## 5. Conclusions

The Cosmic Microwave Background radiation, a highly isotropic remnant of the Recombination Epoch in Big Bang cosmology, has been shown to be a highly applicable Random Bit Generator for asymmetric and symmetric cryptography capable of conforming to the United States government's Federal Information Processing Standard FIPS 140-2. When applied to a Vernam cipher and extended to key spaces in excess of the 256 ASCII character set, the CMB power spectrum offers the capacity for a key space that is too large for not only brute force attacks, but also can be too large for the encryption/decryption capacities of present computer systems.